\documentclass[fleqn,12pt,twoside]{article}
\usepackage[headings]{espcrc1}

\readRCS
$Id: espcrc1.tex,v 1.2 2004/02/24 11:22:11 spepping Exp $
\usepackage{graphicx}
\usepackage[figuresright]{rotating}

\newcommand{\AmS}{{\protect\the\textfont2
  A\kern-.1667em\lower.5ex\hbox{M}\kern-.125emS}}
\newcommand{\be}{\begin{eqnarray}}
\newcommand{\ee}{\end{eqnarray}}

\title{The thermal nature of high $p_T$ photons in high energy nuclear collisions}

\author{Charles Gale\address{Department of Physics, McGill University \\ 3600 University Street,  Montreal, QC H3A 2T8 Canada}}%
               
\runtitle{The thermal nature of high $p_T$ photons}
\runauthor{Charles Gale}

\begin{document}

\maketitle

\begin{abstract}
We discuss the recent status of some penetrating electromagnetic probes of relativistic nuclear collisions, and the information contained in their measurement. We concentrate in turn on sources that produce high $p_T$ photons: those of purely thermal origin, those producing direct photons, those related to jet fragmentation, and those associated with the interaction of jets with the colored plasma. Whenever possible, we compare with RHIC data and in some cases we make predictions for the LHC.  
\end{abstract}

\section{INTRODUCTION}
Penetrating probes constitute a key aspect of the entire relativistic heavy ion program. Of these, electromagnetic observables enjoy a priviledged status as their final-state interactions are minimal, owing to the relative small size of the electromagnetic fine structure constant. From the point of view of new physics discoveries, the interesting sources of photons are those where the quark gluon plasma is directly involved in the emission process. However, it is nevertheless crucial to precisely compute the electromagnetic emissivity of all phases, as the hot matter will emit throughout its history. Quite generally, the emission rate, $R$, for real photons and leptons pairs can be written in terms of the in-medium, retarded photon self-energy at finite temperature \cite{kg}:
\begin{eqnarray}
k_0 \frac{d^3 R}{d^3 k} = - \frac{g^{\mu \nu}}{(2 \pi)^3} {\rm Im} \Pi^{\rm R}_{\mu \nu} (k) \frac{1}{e^{\beta k_0 } - 1} {\parbox{30ex}{\ }\rm (photons)} \nonumber
\end{eqnarray}
\begin{eqnarray}
E_+ E_- \frac{d^6 R}{d^3 p_+ d^3 p_-} = \frac{2 e^2}{(2 \pi)^6} \frac{1}{k^4} L^{\mu \nu} {\rm Im} \Pi^{\rm R}_{\mu \nu} (k) \frac{1}{{\rm e}^{\beta k_0} - 1}\parbox{16ex}{\ } {\rm (dileptons)}
\end{eqnarray}
In the above equation the lepton tensor is simply defined in terms of the lepton four-momenta and mass: $L^{\mu \nu} = p^\mu_+ p^\nu_- + p^\mu_- p^\nu_+ - g^{\mu \nu} \left( p_+ \cdot p_- + m^2 \right)$, and $k^\mu$ is the photon four-momentum (real or virtual). A large effort has been devoted to the evaluation of the thermal emission profile of both hadronic and partonic matter \cite{Gale:2003iz}. The temperature fixes the scale of the emission process, and therefore thermal emission will mostly populate the low- to intermediate-momentum  states of the photon distribution function. The calculation of the electromagnetic emissivity of strongly interacting thermal matter is a topic of considerable interest, that has been boosted recently by the publication of precise dilepton data from the NA60 collaboration at the CERN SPS \cite{Arnaldi:2006jq}. These have provoked a large amount of theoretical activity, which however won't be discussed here owing to space constraints. Since this contribution is mostly about the future, we shall concentrate on photons not of the ``classic'' thermal type \cite{fein,shu}, but on newer sources characterized by the interaction of a hard non-thermal parton with softer thermal components.

\section{JET-PLASMA INTERACTIONS}

In the current era of RHIC experiments, certainly one of the most striking experimental discoveries has been the impressive suppression of single particle yields in nucleus-nucleus collisions, relative to proton-proton collisions. The most popular theoretical interpretation of these experimental findings is one where the hard QCD jet looses a large amount of energy in  a strongly interacting medium  whose degrees of freedom are not those of hadronic matter in the confined sector of the theory \cite{Gyulassy:2003mc}. The dominant mechanism for energy loss is believed to be related to in-medium gluon emission\footnote{However, recent estimates suggest that elastic collisions might also be efficient in dissipating the initial-state energy \cite{peshier}.}.  The importance of jet-plasma interactions should also manifest itself in electromagnetic observables: a hard jet could interact with an in-medium parton and produce a (real or virtual) photon in the final state. A first estimate confirmed the phenomenological importance of the conversion of a leading parton to a photon in the plasma \cite{fms}. Importantly, the assessment of the genuine importance of this signal requires taking into account the jet energy loss, and the in-medium bremsstrahlung radiation. An approach that is capable of consistently incorporating these is that of AMY \cite{amy}. This is the one we adopt here.

\subsection{Hadronic spectra}

\begin{figure}[h!]
\begin{center}
\vspace*{-0.8cm}
\includegraphics*[width=8cm]{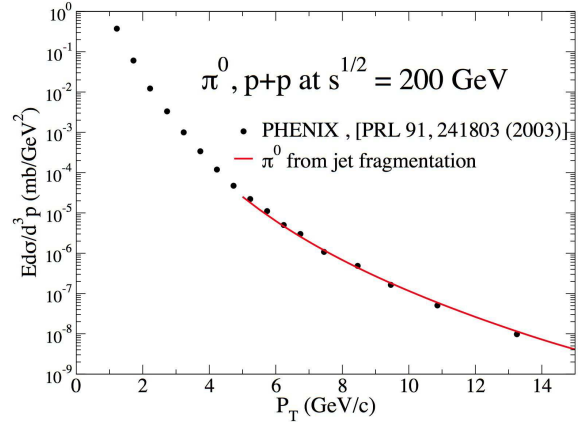}
\vspace*{-1.cm}
\caption{\small Neutral pion spectrum in $pp$ collisions at RHIC. The data are from PHENIX \protect\cite{phenix_pions}, and the solid line is the calculated result from jet fragmentation \protect\cite{tgjm}}

\label{pion}
\end{center}
\end{figure}

A preliminary but necessary step consists of verifying the adequacy of the model with hadronic measurements.  A baseline calculation of the neutral pion spectrum in nucleon-nucleon collisions is given by
\begin{eqnarray}
E_\pi \frac{d^3 \sigma_{\rm NN}}{d^3 p_\pi} = \sum_{a, b, c, d} \int d x_a d x_b g(x_a, Q) g(x_b, Q) {\rm K_{jet}} \frac{d \sigma^{a + b \to c +d}}{d t} \frac{1}{\pi z} D_{\pi^0 /c} (z, Q')
\end{eqnarray}
where $g(x, Q)$ is the parton distribution function (PDF) in a nucleon, $D_{\pi^0 /c} (z, Q')$ is the pion fragmentation function, $\frac{d \sigma^{a+b \to c+d}}{d t}$ is the leading order parton-parton cross section, and ${\rm K_{jet}}$ accounts for higher order effects (here ``jet'' is defined as a fast parton with $p_T^{\rm jet} \gg$ 1 GeV. This correction factor is believed not to be strongly dependent on  transverse momentum \cite{Barnafoldi:2000dy}. Using the results of this investigation, we use ${\rm K_{jet}} \sim 1.7$ for RHIC and 1.6 for the LHC. Also, we set the factorization scale ($Q$) and the fragmentation scae $Q'$ both equal to $p_T$. The CTEQ5 parton distribution functions are used, together with fragmentation functions extracted from the results of $e^+ e^-$ collisions \cite{bkk}. Figure \ref{pion} shows the result of this calculation of the high $p_T$ spectrum of neutral pions with the experimental results of the PHENIX collaboration \cite{phenix_pions}. 
\begin{figure}[t!]
\begin{center}
\vspace*{-0.8cm}
\includegraphics*[width=10cm]{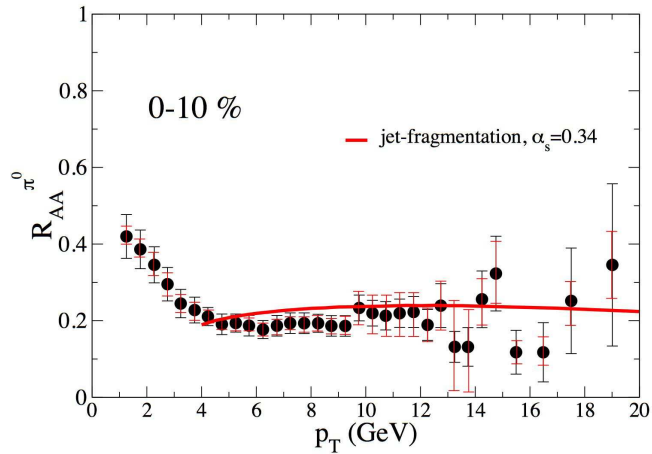}
\vspace*{-1cm}
\caption{\small Nuclear modification factor for neutral pions produced in symmetric collisions of gold nuclei at 200 GeV/nucleon. The data are from the PHENIX collaboration \protect\cite{bc}, the solid red line is the result of the calculation described in the text and references. The black error bars represent statistical errors, while the red error bars represent point-to-point systematic errors.}
\vspace*{-1cm}
\label{RAA}
\end{center}
\end{figure}
To calculate the high $p_T$ spectrum in nucleus-nucleus collisions, this calculation is modified in two ways. First, the PDFs in a nucleus are different from what they are in a nucleon \cite{eks}. Second, we must account for the energy loss of the parton between its production in the initial hard scattering event and its hadronization. We shall pose that a jet fragments outside the medium, as may be justified by estimating the formation time of a typical-energy pion. The fragmentation then proceeds in vacuum, as it does in $e^+ e^-$ collisions,  but with a reduced energy. The jet quenching characteristics of AMY are discussed in \cite{jm}, and the details of our space-time modeling of the nuclear collision are discussed in \cite{tgjm} \footnote{Our current analyses rely on a three-dimensional hydrodynamic model: results are forthcoming}. In AMY, the parton energy profile is time-evolved through a set of coupled Fokker-Planck equations \cite{tgjm} for the parton probability distribution, 
where the transition rates $d\Gamma/dk dt$ for various processes contain a consistent handling (up to leading order in $\alpha_s$) of the enhancement provided by co-linear singularities \cite{amy}. 

In this approach, an overall parameter is the strength of the strong coupling constant, as monitored by $\alpha_s$. One may fix this by considering the measured value of the nuclear modification factor, $R_{AA}$:
\begin{eqnarray}
R_{AA} (p_T ) = \frac{d^2 N^{AA} / d p_T d y}{<N_{\rm coll}> d^2 \sigma^{pp}/ d p_T d y} \sigma^{pp}_{\rm inelastic}
\end{eqnarray}
In Figure \ref{RAA}, the result of the nuclear modification factor for $\pi^0$ in central (0 - 10\%) collisions of gold nuclei at 200 GeV/nucleon is shown, together with our theoretical results \cite{bc}. 
We use here $\alpha_s$ = 0.34. 
\begin{figure}[hb!]
\begin{center}
\vspace*{-1.cm}
\includegraphics*[width=10cm]{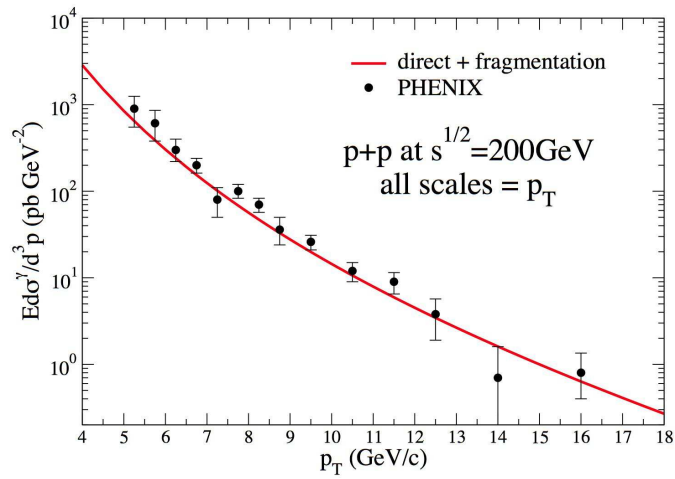}
\vspace*{-1.cm}
\caption{\small The spectrum of prompt photons produced in $pp$ collisions at RHIC. The data points are from \protect\cite{Reygers:2005sm}, the solid is calculated with Eq. (\ref{prompt}).}
\vspace*{-1.2cm}
\label{gamma_pp}
\end{center}
\end{figure}

\subsection{Photon production}

\begin{figure}
\begin{center}
\vspace*{-.8cm}
\includegraphics*[width=3in]{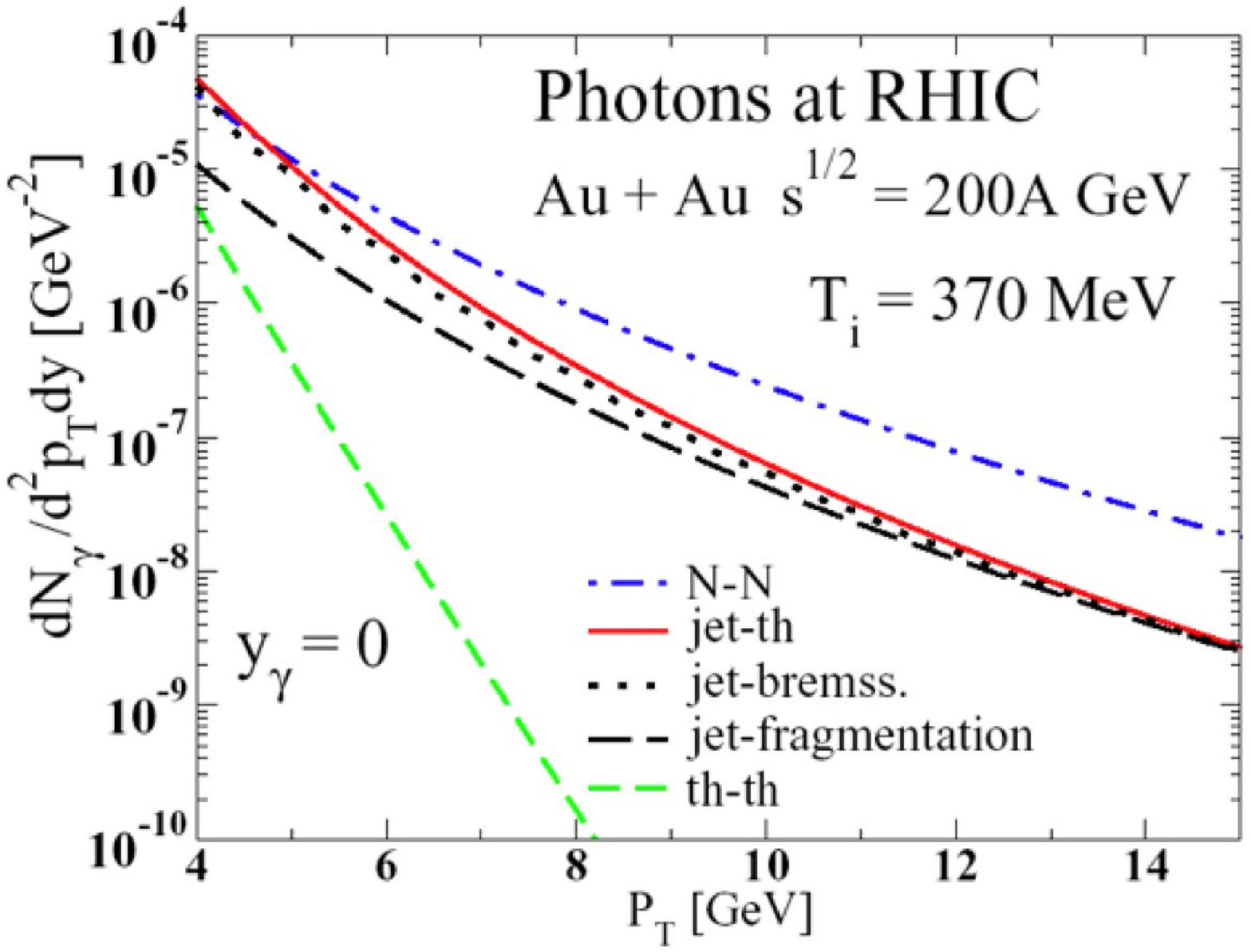}
\includegraphics*[width=3.1in]{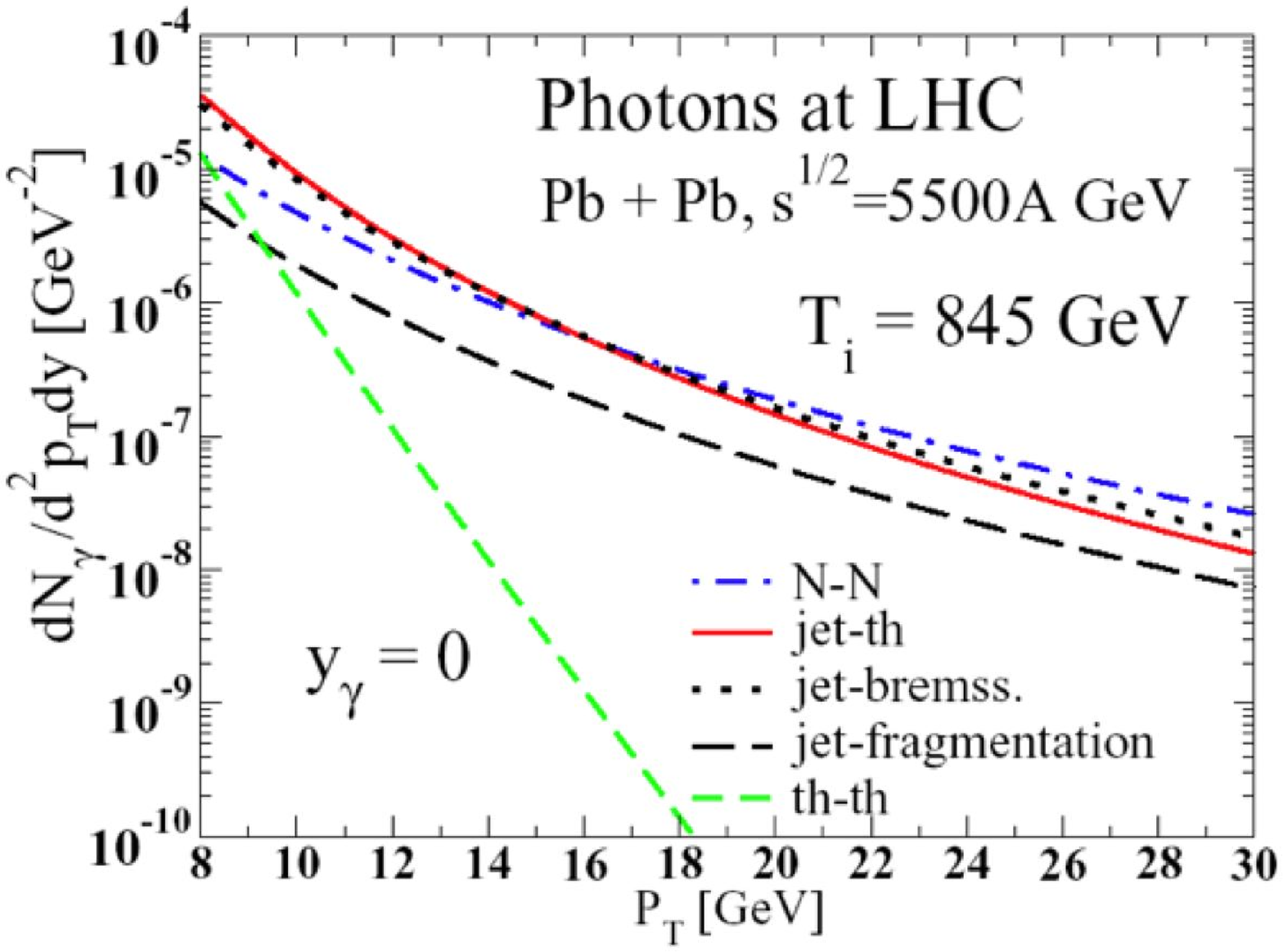}
\vspace*{-1cm}
\caption{\small The left (right) panel shows a decomposition of the spectrum of real photons obtained in Au + Au (Pb + Pb) collisions at RHIC (LHC) energies into its different components. These are photons from Compton and annihilation processes (N-N), jet-thermal and jet-bremsstrahlung photons, those coming from the fragmentation of  escaping jets, and photons from the thermal radiation of the cooling system. The parameter values (temperature, etc \ldots) and the time-evolution of the system are discussed in \protect\cite{tgjm}, see also \cite{new}.}
\vspace*{-1.cm}
\label{all}
\end{center}
\end{figure}

Having fixed the space-time evolution and the value of the strong coupling, we now turn to the emission of electromagnetic radiation. Here also, a baseline calculation in the case of nucleon-nucleon collisions is necessary. The hard photons produced in these events can be divided in three categories: direct photons, fragmentation photons and background photons. Direct photons are those produced by Compton scattering and annihilation of two incoming partons. Fragmentation photons are those produced by final states partons. Background photons are those produced by the decay of hadrons subsequent to the collision, mainly from $\pi^0 \to \gamma \gamma$. The ``prompt photons'' are the sum of those produced in direct and fragmentation processes. The prompt photon spectrum is given by
\be
\label{pp_prompt}
E_{\gamma}\frac{d\sigma}{d^3p_{\gamma}}&=&\sum_{a,b}\int dx_a dx_b
g(x_a,Q)g(x_b,Q) \Big\{{\rm K_{\gamma}}(p_T)\frac{d\sigma^{a+b\rightarrow
\gamma+d}}{dt}\frac{2x_ax_b}{\pi
(2x_a-2\frac{p_T}{\sqrt{s}}e^y)} \nonumber \\
&&\times 
\delta \left(x_b - \frac{x_a p_T e^{-y}}{x_a \sqrt{s} - p_T
e^y}\right)
+{\rm K_{brem}}(p_T)\frac{d\sigma^{a+b\rightarrow
c+d}}{dt}\frac{1}{\pi z} D_{\gamma/c}(z, Q)\Big\} \, .
\label{prompt}
\ee
where ${\rm K_{\gamma}}$ and ${\rm K_{brem}}$ are correction factors to take into account NLO effects; we
evaluate them using the numerical program from Aurenche 
{\it et al.}~\cite{aurenche}, 
obtaining ${\rm K_{\gamma}}$(10 GeV)$\sim$ 1.5 for RHIC and LHC and ${\rm K_{brem}}$(10 GeV)$\sim$ 1.8 at RHIC and 1.4 at LHC.  All scales (renormalization, factorization and fragmentation) have been set equal to the photon transverse momentum $p_T$.
 As for pions, the photon fragmentation function $D_{\gamma/c}$ is
extracted without medium effects in $e^- e^+$
collisions.
The validity of this treatment
for $pp$ collisions at $\sqrt{s}=$ 200 GeV
is shown in Figure \ref{gamma_pp} with data from PHENIX \cite{Reygers:2005sm}. 
This suggests that the baseline mechanism of high $p_T$ photon production
in nucleon-nucleon collisions at these energies is under quantitative control.
In nucleus-nucleus collisions, additional sources of high $p_T$ photons appear: the medium contributions. These include the direct conversion of a high energy parton to a high energy photon via annihilation with a thermal parton, the in-medium bremsstrahlung of a jet, and the thermal radiation from the plasma. Including all the sources discussed so far yields the summary contained in Figure \ref{all}, for RHIC and the LHC.  
\begin{figure}[h!]
\begin{center}
\vspace*{-.8cm}
\includegraphics*[width=7cm]{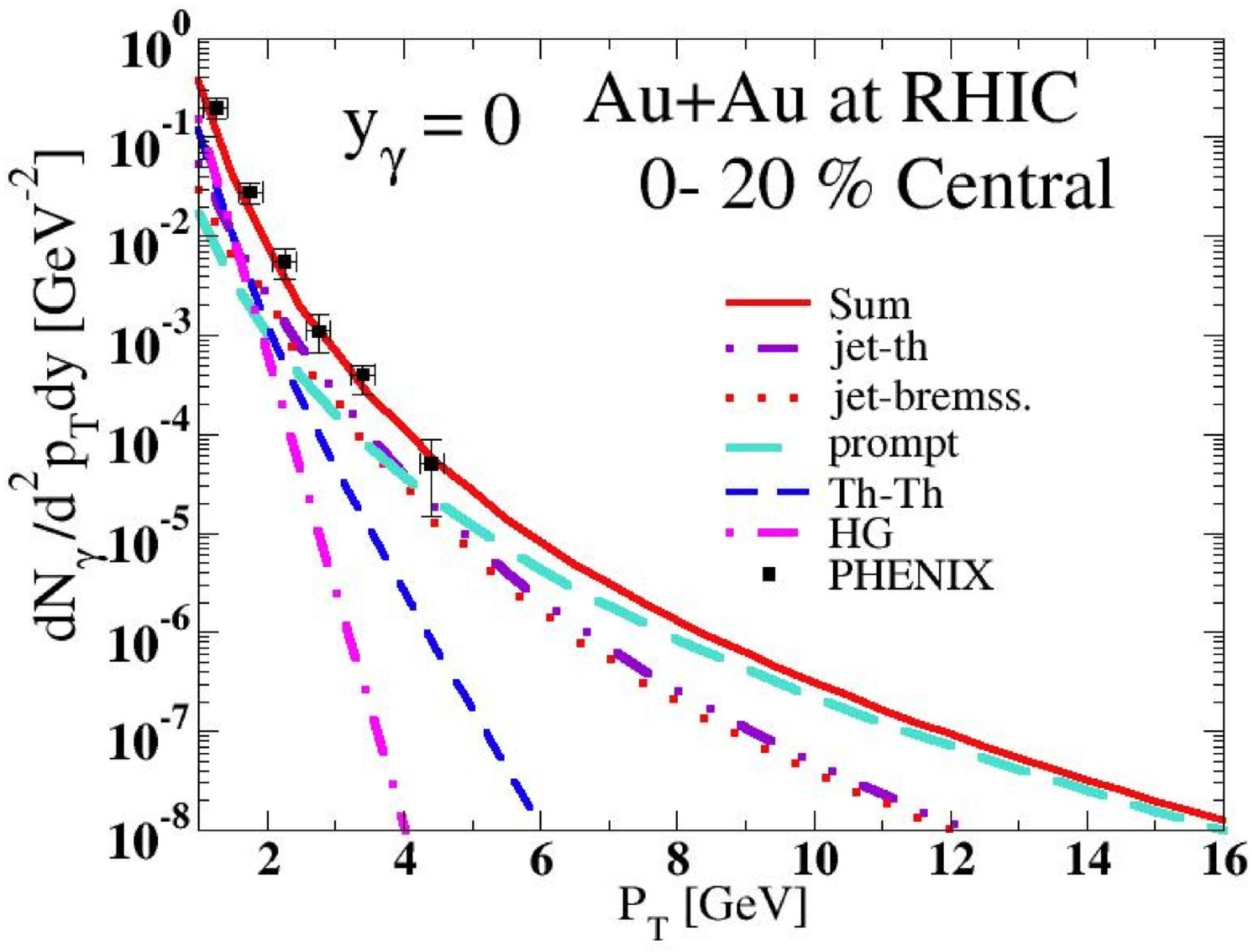}
\includegraphics*[width=7.3cm]{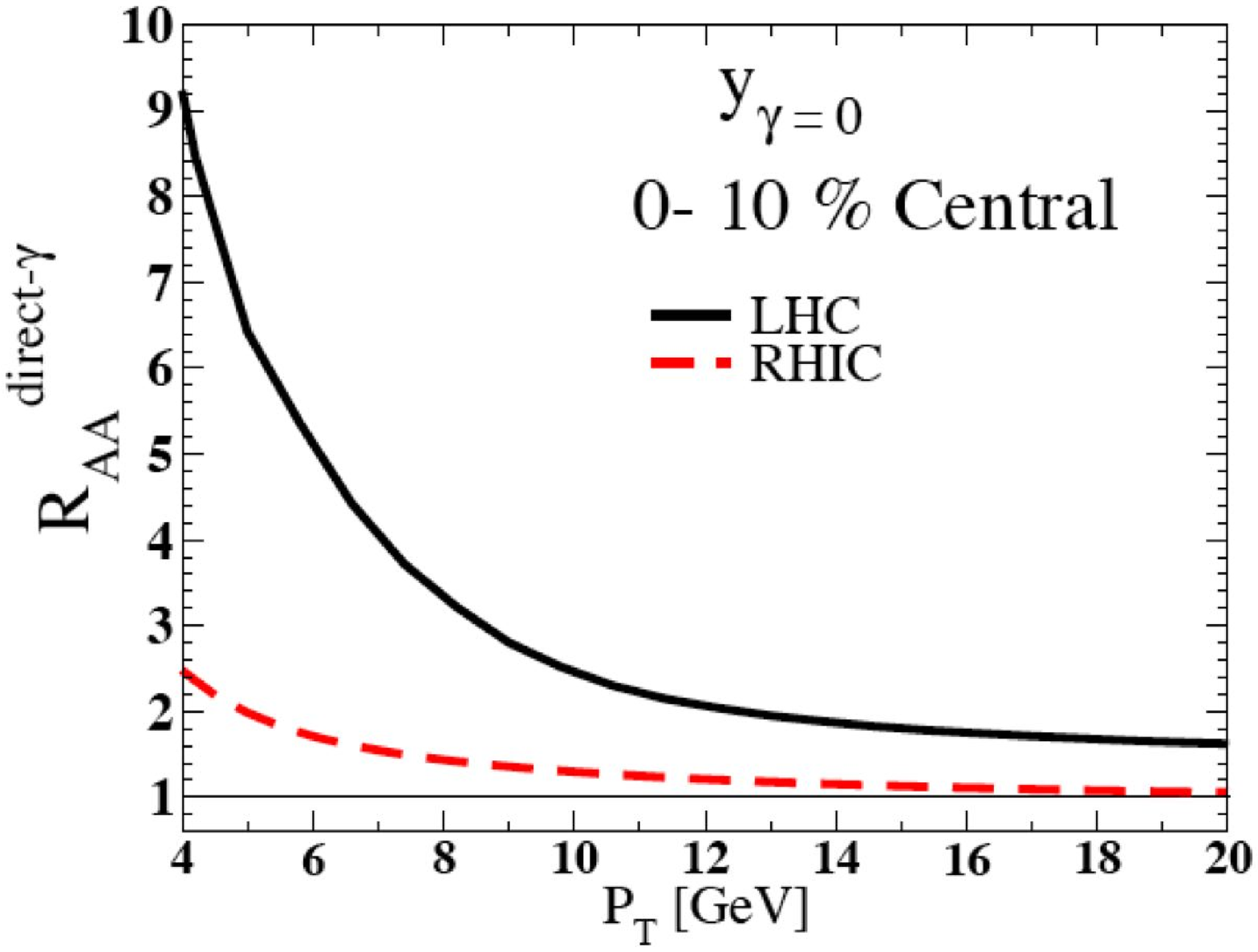}
\vspace*{-1cm}
\caption{\small Left panel: The total photon spectrum (solid curve) produced in central collisions of Au nuclei at RHIC  is decomposed in its different components. Shown here are the contributions associated with Compton, annihilation, and jet-fragmentation (prompt),  with jet-photon conversions (jet-th), with the in-medium bremsstrahlung of jets (jet-bremss), with the thermal radiation from the quark-gluon plasma (Th-Th) \protect\cite{amy}, and with the thermal radiation from the hadronic sector (HG) \protect\cite{TRG}. Right panel: The nuclear modification factor for photons at RHIC (Au + Au) and at the LHC (Pb + Pb). }
\vspace*{-1cm}
\label{all2}
\end{center}
\end{figure}
Another representation of this result at RHIC is given in the left panel of Figure \ref{all2}, where the different contributions are now summed up, and compared with a recent analysis of PHENIX photon data \cite{buesch}.
Interestingly, the curves on this plot constitute a prediction, and were done prior to the publication of the heavy ion photon data. In this theoretical analysis the presence of jet-photon conversions reveal the jet-plasma interactions and thus the presence of conditions suitable for jet-quenching to take place. An alternative view of this physics is given by considering the familiar nuclear modification factor $R_{AA}$ again, but only this time for photons. This is done in the right panel of Figure \ref{all2}: clearly what was a suppression for pions is now an enhancement.

These electromagnetic data represent measurements that complement that of the single hadronic spectrum suppression. However, as it is often the case in the complicated environment associated with relativistic nuclear collisions, a genuine doubt concerning the possibility of precisely determining the initial temperature through measurements like these may persist. When the presence (or absence) of a specific channel is related to an overall normalization, caution must be exerted. 

\begin{figure}[h!]
\begin{center}
\includegraphics*[width=7.6cm]{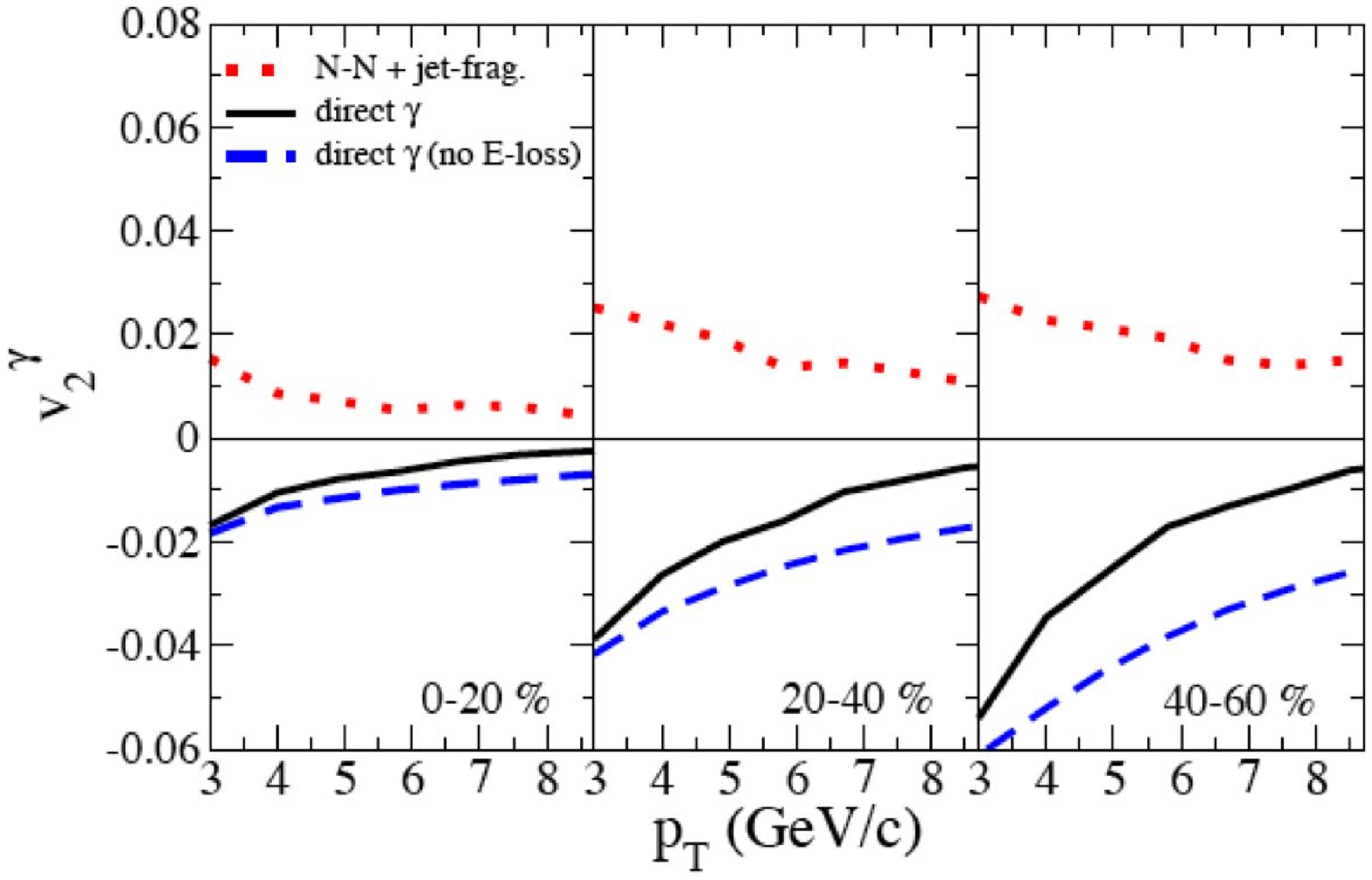}
\includegraphics*[width=7.9cm]{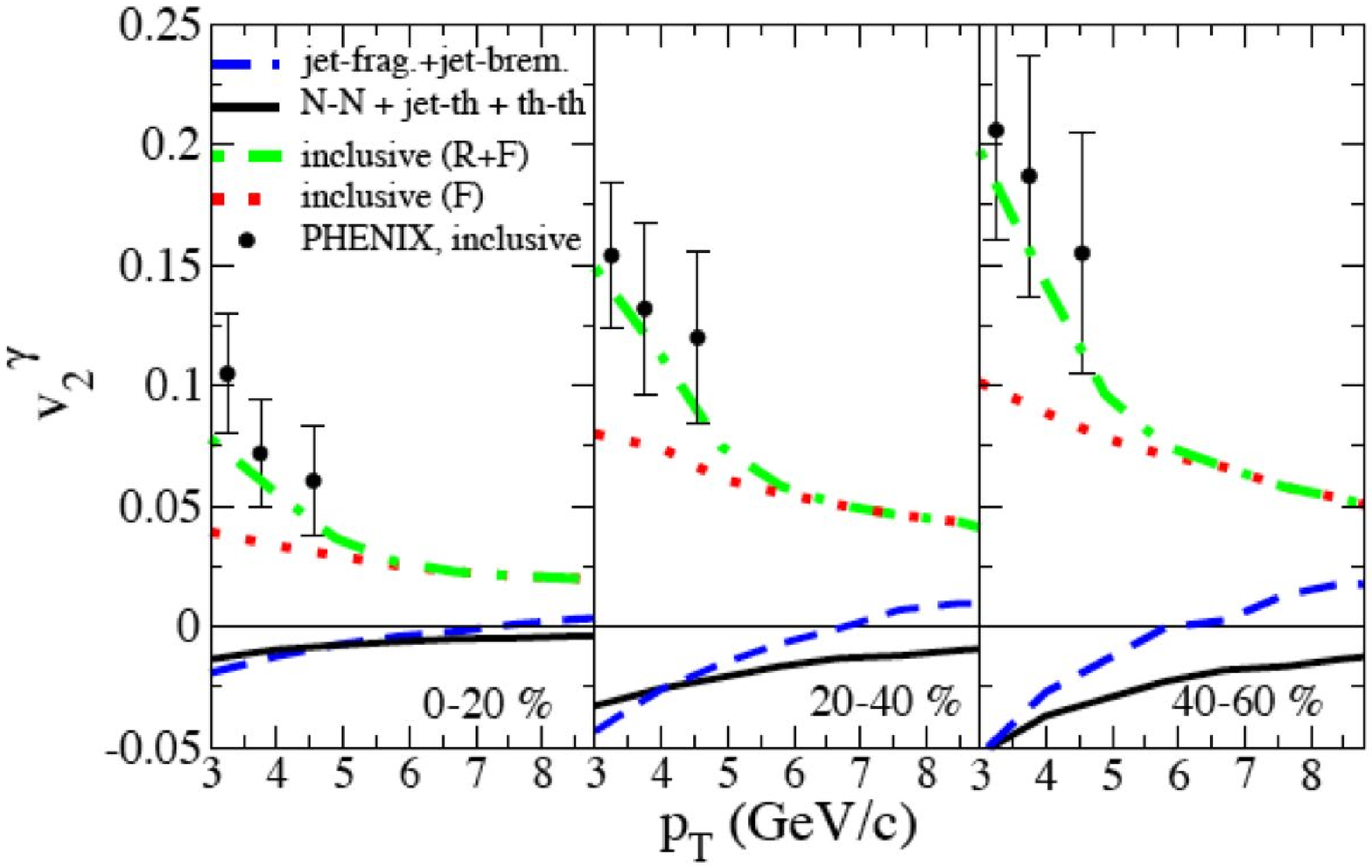}
\vspace*{-1cm}
\caption{\small Left panel: Photon $v_2$ as a function of $p_T$ for Au + Au collisions at RHIC, for three centrality bins. The dotted is the $v_2$ for primary hard photons and jet fragmentation, the solid line is the result with all direct photons. Energy loss is included in both cases. The dashed line is the result without energy loss. Right panel: The dashed line shows jet-fragmentation and induced bremsstrahlung only while the solid curve is associated with jet-photon conversion, primary hard and thermal photons. The dotted curve shows direct photons and the background associated with the decay of neutral mesons coming from jets. The dot-dashed line adds photons from the decay of recombined pions. The data is from PHENIX \protect\cite{Adler:2005rg}.}
\vspace*{-1cm}
\label{v2}
\end{center}
\end{figure}
\subsection{Azimuthal asymmetry of the electromagnetic signal}
To circumvent the possible ambiguity related to simple signal strength in the identification of new physics, an experimental observable which represents a different approach is needed. In this context, it was recently proposed that a sizeable azimuthal asymmetry could be expected for large $p_T$ direct photons produced in non-central high energy nuclear collisions \cite{TGF}, see also \cite{new}. As usual, the anisotropy can be characterized in terms of the Fourier coefficients $v_k$ defined from the differential particle yield through
\be
\frac{dN}{p_Tdp_Td\phi} = \frac{dN}{2\pi p_Tdp_T} \left[ 1+
\sum_k 2 v_k(p_T) \cos( k\phi) \right]
\ee
where the angle $\phi$ is defined with respect to the reaction plane. The physics in this measurement can be summarized simply: at RHIC all measured hadronic species have a positive $v_2$, which corresponds to a larger yield in  direction parallel to the short axis of the ellipse defined by the initial nuclear geometry. However, jet-plasma photons will be preferentially emitted along the large axis of the ellipse, as that probability grows with the path traversed in the medium. Using the same technology introduced earlier, we carry out a calculation for Au + Au collisions at RHIC energies, for three classes of centrality. The left panel of Figure \ref{v2} shows the photon $v_2$ as a function of $p_T$. The dotted line is the result for primary hard direct photons including photons produced from fragmentation. As expected, photons from fragmentation lead to a positive $v_2$ which is then somewhat reduced when adding primary hard photons. The solid line is a result which now includes the addition of bremsstrahlung, jet-photon conversion and thermal photons: the $v_2$ associated with induced bremsstrahlung and jet-photon conversion is indeed negative. Together they are able to overcome the positive $v_2$ from fragmentation, leading to an overall negative elliptic asymmetry for direct photons at moderate $p_T$. Note that moving to even lower momenta would require the inclusion of the asymmetry associated with collective hydrodynamic behaviour \cite{Chatterjee:2005de}. The right panel of the same figure shows what are the experimental expectations. The dotted line shows $v_2$ for inclusive photons before background subtraction. Here, the signal is dominated by the contribution from decays of $\pi^0$ and $\eta$. The resulting $v_2$ is positive and large. Only hadrons from fragmentation have been included. However, it has been proposed that hadron production up to a $p_T$ of 4 to 6 GeV/c can be interpreted in terms of a significant contribution from the recombination of quarks \cite{Fries:2003kq}. The dot-dashed line shows the $v_2$ of inclusive photons if decays of $\pi^0$ and $\eta$ from recombination are included. Data on azimuthal asymmetry of photons have been measured: those are also shown and there is good agreement with our calculations. 
Note that the sources of photons described by the dashed and solid lines actually represent two classes of events: those that can be correlated with a jet, and those independent of hard hadronic particles. This opens the possibility of disentangling the different channels through experimental isolation cuts, for example.

\section{CONCLUSION}
The photons (and dileptons \cite{simon2}) originating from jet-plasma interactions are seen to constitute a source bright enough to be detected experimentally. Furthermore, we have shown that their azimuthal distributions can exhibit a characteristic behaviour. In general, electromagnetic observables will continue to reveal important physics, complementary to that learned in measurements of strongly interacting particles. It is also fair to say that we are witnessing a shift of paradigm in what concerns the calculation of real and virtual photon production. At RHIC energies, where $pp$, $pA$ and $AA$ data now exists, it appears that perturbative QCD (up to next-to-leading order in the strong coupling) offers a reliable quantitative description of the data \cite{Aurenche:1} in the first two cases. One then has confidence in a solid baseline on which to build $AA$ analyses. This fact constitutes a necessary requirement in order to enter an era of precision heavy ion physics: the future is indeed bright!

\section*{Acknowledgments}
I am happy to acknowledge the help of all my collaborators, and in particular that of Simon Turbide, with whom all the work presented here was done. This research was funded in part by the Natural Sciences and Engineering Research Council of Canada, and in part by the Fonds Qu\'eb\'ecois de Recherche sur la Nature et les Technologies.


\begin{thebibliography}{99}
\bibitem{kg}L. D. McLerran and T. Toimela, Phys. Rev. D {\bf 31}, 545 (1985); H. A. Weldon,
Phys. Rev. D {\bf 42}, 2384 (1990); C. Gale and J. I. Kapusta, Nucl. Phys. {\bf B357},
65 (1991).
\bibitem{Gale:2003iz}See, for example, 
  C.~Gale and K.~L.~Haglin,
  arXiv:hep-ph/0306098, and references  therein.
\bibitem{Arnaldi:2006jq}
  R.~Arnaldi {\it et al.}  [NA60 Collaboration],
   ``First measurement of the rho spectral function in high-energy nuclear
  Phys.\ Rev.\ Lett.\  {\bf 96}, 162302 (2006)
  [arXiv:nucl-ex/0605007].  
\bibitem{fein}
  E.~L.~Feinberg,
  Nuovo Cim.\ A {\bf 34}, 391 (1976).
\bibitem{shu}
  E.~V.~Shuryak,
  Phys.\ Lett.\ B {\bf 78}, 150 (1978)
  [Sov.\ J.\ Nucl.\ Phys.\  {\bf 28}, 408.1978\ YAFIA,28,796 (1978\ YAFIA,28,796-808.1978)].
\bibitem{Gyulassy:2003mc} See, for example, 
  M.~Gyulassy, I.~Vitev, X.~N.~Wang and B.~W.~Zhang,
  arXiv:nucl-th/0302077, and references therein.
\bibitem{peshier}
  A.~Peshier,
  arXiv:hep-ph/0607299.
  \bibitem{fms}
  R.~J.~Fries, B.~Muller and D.~K.~Srivastava,
  Phys.\ Rev.\ Lett.\  {\bf 90}, 132301 (2003)
  [arXiv:nucl-th/0208001].
\bibitem{amy}P.~Arnold, G.~D.~Moore and L.~G.~Yaffe,
  JHEP {\bf 0011}, 001 (2000)
  [arXiv:hep-ph/0010177]; 
  P.~Arnold, G.~D.~Moore and L.~G.~Yaffe,
  JHEP {\bf 0112}, 009 (2001)
  [arXiv:hep-ph/0111107]; 
  P.~Arnold, G.~D.~Moore and L.~G.~Yaffe,
  JHEP {\bf 0206}, 030 (2002)
  [arXiv:hep-ph/0204343].

\bibitem{Barnafoldi:2000dy}
  G.~G.~Barnafoldi, G.~I.~Fai, P.~Levai, G.~Papp and Y.~Zhang,
  J.\ Phys.\ G {\bf 27}, 1767 (2001)
  [arXiv:nucl-th/0004066].
\bibitem{bkk}
  J.~Binnewies, B.~A.~Kniehl and G.~Kramer,
  Phys.\ Rev.\ D {\bf 52}, 4947 (1995)
  [arXiv:hep-ph/9503464].
\bibitem{phenix_pions}
S.~S.~Adler {\it et al.}  [PHENIX Collaboration],
  Phys.\ Rev.\ Lett.\  {\bf 91}, 241803 (2003)
  [arXiv:hep-ex/0304038].
\bibitem{eks}
  K.~J.~Eskola, V.~J.~Kolhinen and C.~A.~Salgado,
  Eur.\ Phys.\ J.\ C {\bf 9}, 61 (1999)
  [arXiv:hep-ph/9807297].
  \bibitem{jm}
S.~Jeon and G.~D.~Moore,
  Phys.\ Rev.\ C {\bf 71}, 034901 (2005)
  [arXiv:hep-ph/0309332].
\bibitem{tgjm}
S.~Turbide, C.~Gale, S.~Jeon and G.~D.~Moore,
  Phys.\ Rev.\ C {\bf 72}, 014906 (2005)
  [arXiv:hep-ph/0502248].
\bibitem{bc}
B.~A.~Cole,
  Nucl.\ Phys.\ A {\bf 774} (2006) 225.
\bibitem{aurenche}
P. Aurenche, R. Baier, A. Douiri, M. Fontannaz and D. Schiff, Nucl. Phys. B
{\bf 286}, 553 (1987); Nucl. Phys. {\bf 297}, 661 (1988).
\bibitem{Reygers:2005sm}
  K.~Reygers  [PHENIX Collaboration],
  Eur.\ Phys.\ J.\ C {\bf 43}, 393 (2005)
  [arXiv:nucl-ex/0502018].
\bibitem{buesch}H. Buesching [PHENIX Collaboration], Quark Matter 2005, Budapest, Hungary.
\bibitem{new} Some of the figures in the quoted reference contain a numerical inaccuracy. The ones shown here constitute the corrected versions.
\bibitem{TRG}
S.~Turbide, R.~Rapp and C.~Gale,
  Phys.\ Rev.\ C {\bf 69}, 014903 (2004)
  [arXiv:hep-ph/0308085].
\bibitem{TGF}
S.~Turbide, C.~Gale and R.~J.~Fries,
  Phys.\ Rev.\ Lett.\  {\bf 96}, 032303 (2006)
  [arXiv:hep-ph/0508201].
\bibitem{Chatterjee:2005de}
  R.~Chatterjee, E.~S.~Frodermann, U.~W.~Heinz and D.~K.~Srivastava,
  Phys.\ Rev.\ Lett.\  {\bf 96}, 202302 (2006)
  [arXiv:nucl-th/0511079].  
  \bibitem{Adler:2005rg}
  S.~S.~Adler {\it et al.}  [PHENIX Collaboration],
   ``Measurement of identified pi0 and inclusive photon v(2) and implication  to
  Phys.\ Rev.\ Lett.\  {\bf 96}, 032302 (2006)
  [arXiv:nucl-ex/0508019].
\bibitem{Fries:2003kq}
  R.~J.~Fries, B.~Muller, C.~Nonaka and S.~A.~Bass,
  Phys.\ Rev.\ C {\bf 68}, 044902 (2003)
  [arXiv:nucl-th/0306027].
\bibitem{simon2}
  S.~Turbide, C.~Gale, D.~K.~Srivastava and R.~J.~Fries,
  Phys.\ Rev.\ C {\bf 74}, 014903 (2006)
  [arXiv:hep-ph/0601042].
\bibitem{Aurenche:1}
  P.~Aurenche, J.~P.~Guillet, E.~Pilon, M.~Werlen and M.~Fontannaz,
  Phys.\ Rev.\ D {\bf 73}, 094007 (2006).

\end{thebibliography}
\end{document}